\documentstyle[12pt]{article}

\begin{document}
\setcounter{page}{1}
\pagestyle{plain} \vspace{1cm}
\begin{center}
\Large{\bf On the Quantum-Corrected Black Hole Thermodynamics}\\
\small
\vspace{1cm} {\bf Kourosh Nozari$^{1,2}$}\quad and \quad {\bf S. Hamid Mehdipour$^{1}$}\\
\vspace{0.5cm} {\it $^{1}$Department of Physics,
Faculty of Basic Sciences,\\
University of Mazandaran,\\
P. O. Box 47416-1467,
Babolsar, IRAN\\
e-mail: knozari@umz.ac.ir}

\end{center}
\vspace{1.5cm}

\begin{abstract}
Bekenstein-Hawking Black hole thermodynamics should be corrected to
incorporate quantum gravitational effects. Generalized Uncertainty
Principle(GUP) provides a perturbational framework to perform such
modifications. In this paper we consider the most general form of
GUP to find black holes thermodynamics in microcanonical ensemble.
Our calculation shows that there is no logarithmic pre-factor in
perturbational expansion of entropy. This feature will solve part of
controversies in
literatures regarding existence or vanishing of this pre-factor.
\end{abstract}
\newpage

\section{Introduction}
Quantum geometry, string theory and loop quantum gravity all
indicate that measurements in quantum gravity should be governed by
generalized uncertainty principle[1-5]. As a result, there is a
minimal length scale of the order of Planck length which can not be
probed. In the language of string theory, this is related to the
fact that a string can not probe distances smaller than its length.
Therefore, it seems that a re-formulation of quantum theory to
incorporate gravitational effects from very beginning is necessary
to investigate Planck scale physics. Introduction of this idea, has
drawn considerable attentions and many authors have considered
various problems in the framework of generalized uncertainty
principle[6-20]. Quantum gravitational induced corrections to black
hole thermodynamics as a consequence of GUP are studied with details
in literatures. Adler and his coworkers[21] have argued that
contrary to standard viewpoint, GUP may prevent small black holes
total evaporation in exactly the same manner that the ordinary
uncertainty principle prevents the Hydrogen atom from total
collapse. They have considered these black holes remnants as a
possible source of dark matter. Medved and Vagenas[22], have
recently formulated the quantum corrected entropy of black holes in
terms of an expansion and have claimed that this expansion is
consistent with all previous findings. Bolen and Cavaglia, have
obtained thermodynamical properties of Schwarzschild anti-de Sitter
black holes using GUP [23]. They have considered two limits of their
equations, quantum gravity limit and usual quantum mechanical regime
and in each circumstances they have interpreted their results.
Action for the exact string black hole has been considered by
Grumiller and he has found exact relation for entropy of a string
black hole[24]. Existence or vanishing of logarithmic prefactor in
the expansion of black hole entropy has been considered in details
by Medved. He has argued in [25] that "the best guess for the
prefactor might simply be zero" regarding to mutual cancelation of
canonical and microcanonical contributions. But later, considering
some general considerations of ensemble theory, he has argued that
canonical and microcanonical corrections could not cancel each other
to result in vanishing logarithmic pre-factor in entropy[26].
Meanwhile, Hod has employed statistical arguments that constrains
this prefactor to be a non-negative integer[27]. There are other
literatures considering logarithmic corrections to black hole
entropy[28,29], but there is no explicit statement about the
ultimate value of this prefactor. \\
Here, using generalized uncertainty principle in its most general
form as our primary input, we find explicit perturbational expansion
of black hole entropy in microcanonical ensemble. By computing the
coefficients of this expansion, we will show that there is no
logarithmic prefactor in expansion of microcanonical entropy.

\section{ Generalized Uncertainty Principle }
Usual uncertainty principle of quantum mechanics, the so-called
Heisenberg uncertainty principle, should be re-formulated regarding
to non-commutative nature of spacetime. It has been indicated that
in quantum gravity there exists a minimal observable distance on the
order of the Planck length which in the context of string theories,
this observable distance is referred to GUP[1-5],[30-33]. A
generalized uncertainty principle can be formulated as
\begin{equation}
\label{math:2.1} \delta x\geq \frac{\hbar}{2\delta p} + const.
G\delta p,
\end{equation}
which, using string theoretical arguments regarding the minimal
nature of $l_p$[4], can be written as
\begin{equation}
\label{math:2.2} \delta x\geq \frac{\hbar}{2\delta p} +
\alpha^{2}l_{p}^2\frac{\delta p}{2\hbar}
\end{equation}
The corresponding Heisenberg commutator now becomes,
\begin{equation}
\label{math:2.3} [x,p]=i\hbar(1+\alpha' p^2).
\end{equation}
Note that commutator (3) is not the direct consequence of relation
(2), but can be considered as one of its consequences[11]. $\alpha$
is positive and independent of $\delta x$ and $\delta p$ but may in
general depend to the expectation values of $x$ and $p$. In the same
manner one can consider the following generalization,
\begin{equation}
\label{math:2.4} \delta x\delta p\geq
\frac{\hbar}{2}\Big(1+\frac{\beta^{2}}{l_{p}^{2}}(\delta x)^2\Big),
\end{equation}
which indicates the existence of a minimal observable momentum. It
is important to note that GUP itself can be considered as a
perturbational expansion[11]. In this viewpoint, one can consider a
more general statement of GUP as follows
\begin{equation}
\label{math:2.5} \delta x\delta p\geq
\frac{\hbar}{2}\bigg(1+\frac{{\alpha}^2 l^2_p}{\hbar^2}(\delta
p)^2+\frac{{\beta}^2}{l^2_p}(\delta x)^2+\gamma\bigg),
\end{equation}
where $\alpha$, $\beta$ and $\gamma$ are positive and independent of
$\delta x$ and $\delta p$ but may in general depend to the
expectation values of $x$ and $p$. Here, Planck length is defined as
$ l_{p}=\sqrt{ \frac{\hbar G}{c^{3}}}$. Note that (5) leads to
nonzero minimal uncertainty in both position $(\delta x)_{min}$ and
momentum $(\delta p)_{min}$. In which follows, we use this more
general form of GUP as our primary input and construct a
perturbational framework to find thermodynamical properties of black
hole and their quantum gravitational corrections. It should be noted
that since GUP is a model independent concept[6], the results which
we obtain are consistent with any promising theory of quantum
gravity.

\section{Black Holes Thermodynamics}
Consider the most general form of GUP as equation (5). A simple
calculation gives,
\begin{equation}
\label{math:3.1} \delta x\simeq\frac{l_p^2 \delta p}{{\beta}^2
\hbar}\Bigg[1\pm\sqrt{1-{\beta}^2\bigg(
{\alpha}^2+\frac{\hbar^2(\gamma+1)}{l_p^2(\delta
p)^2}\bigg)}\,\,\Bigg].
\end{equation}
Here, to achieve standard values (for example $\delta x\delta p\geq
\hbar$) in the limit of $\alpha=\beta=\gamma =0$, we should consider
the minus sign. One can minimize $\delta x$ to find
\begin{equation}
\label{math:3.2}(\delta x)_{min}\simeq\pm\alpha
l_p\sqrt{\frac{(1+\gamma)}{1-\alpha^2\beta^2}}.
\end{equation}
The minus sign, evidently has no physical meaning for minimum of
position uncertainty. Therefore, we find
\begin{equation}
\label{math:3.2}(\delta x)_{min}\simeq \alpha
l_p\sqrt{\frac{(1+\gamma)}{1-\alpha^2\beta^2}}.
\end{equation}
This equation gives the minimal observable length on the order of
Planck length. Since in our definition, $\alpha$ and $\beta$ are
dimensionless positive constant always less than one(extreme quantum
gravity limit), $(\delta x)_{min}$ is defined properly. Equation (5)
gives also
\begin{equation}
\label{math:3.3} \delta p\simeq\frac{\hbar \delta
x}{{\alpha}^2l_p^2}\Bigg[1\pm\sqrt{1-{\alpha}^2\bigg({\beta}^2+\frac{
l_p^2(\gamma+1)}{(\delta x)^2}\bigg)}\Bigg].
\end{equation}
Here to achieve correct limiting results we should consider the
minus sign in round bracket. From a heuristic argument based on
Heisenberg uncertainty relation, one deduces the following equation
for Hawking temperature of black holes[21],
\begin{equation}
\label{math:3.4}T_H\approx \frac{\delta p c}{2\pi}
\end{equation}
Based on this viewpoint, but now using generalized uncertainty
principle in its most general form, modified black hole temperature
in GUP is,
\begin{equation}
\label{math:3.5}T^{GUP}_{H}\approx \frac{\hbar c \delta x}{{2\pi
\alpha}^2 l^2_p}\Bigg[1-\sqrt{1-{\alpha}^2\bigg(
{\beta}^2+\frac{l^2_p(\gamma+1)}{(\delta x)^2}\bigg)}\,\,\Bigg].
\end{equation}
Now consider a quantum particle that starts out in the vicinity of
an event horizon and then ultimately absorbed by black hole. For a
black hole absorbing such a particle with energy $E$ and size $R$,
the minimal increase in the horizon area can be expressed as [34]
\begin{equation}
\label{math:3.6}(\Delta A)_{min}\geq \frac{8\pi l_p^2ER}{\hbar c},
\end{equation}
then one can write
\begin{equation}
\label{math:3.7}(\Delta A)_{min}\geq\frac{8\pi l_p^2\delta p c\delta
x}{\hbar c},
\end{equation}
where $E\sim c\delta p $ and  $R\sim\delta x$. Using equation
(9)(with minus sign) for $\delta p$ and defining $A=4\pi
(\frac{\delta x_{min}}{2})^{2}$, we find
\begin{equation}
\label{math:3.8} (\Delta A)_{min}\simeq\frac{8
A}{{\alpha}^2}\Bigg[1-\sqrt{1-{\alpha}^2\bigg({\beta}^2+\frac{\pi
l_p^2(\gamma+1)}{A}\bigg)}\Bigg].
\end{equation}
Now we should determine $\delta x$. Since our goal is to compute
microcanonical entropy of a large black hole, near-horizon geometry
considerations suggests the use of inverse surface gravity or simply
twice the Schwarzschild radius for $\delta x$. Therefore, $\delta
x\approx 2r_s$ and defining $4\pi r_s^2=A$ and $(\Delta
S)_{min}=b=constant$, then it is easy to show that,
\begin{equation}
\label{math:3.9}\frac{dS}{dA}\simeq\frac{(\Delta S)_{min}}{(\Delta
A)_{min}}\simeq \frac{b{\alpha}^2}{8
A\Bigg[1-\sqrt{1-{\alpha}^2\bigg({\beta}^2+\frac{\pi
l_p^2(\gamma+1)}{A}\bigg)}\Bigg]}.
\end{equation}
Three point should be considered here. First note that $b$ can be
considered as one bit of information since entropy is an extensive
quantity. Considering "calibration factor" of Bekenstein as
$\frac{ln 2}{2\pi}$, the minimum increase of entropy(i.e. $b$),
should be $ln 2$. Secondly, note that
$\frac{dS}{dA}\simeq\frac{(\Delta S)_{min}}{(\Delta A)_{min}}$ holds
since this is an approximate relation and give only relative changes
of corresponding quantities. As the third remarks, our approach
considers microcanonical ensemble since we are dealing with a
Schwarzschild black hole of fixed mass. Now we should perform
integration. There are two possible choices for lower limit of
integration, $A=0$ and $A=A_p$ . Existence of a minimal observable
length leads to existence of a minimum event horizon area, $A_p =
4\pi \Big(\frac{(\delta x)_{min}}{2}\Big)^{2}$. So it is physically
reasonable to set $A_p$ as lower limit of integration. This is in
accordance with existing picture[21]. Based on these arguments, we
can write
\begin{equation}
\label{math:3.10} S\simeq\int_{A_p}^A\frac{b{\alpha}^2}{8
A\Bigg[1-\sqrt{1-{\alpha}^2\bigg({\beta}^2+\frac{\pi
l_p^2(\gamma+1)}{A}\bigg)}\Bigg]}dA.
\end{equation}
Integration gives,
\begin{equation}
\label{math:3.11} S\simeq\mu\Bigg[\ln\Bigg|\frac{-2\sqrt{A(\zeta A
+\eta)}+ A(\zeta +1)+\eta}{-2\sqrt{A_p(\zeta A_p +\eta)}+ A_p(\zeta
+1)+\eta}\Bigg| +\sqrt{\zeta}\ln\Bigg|\frac{\eta+2\zeta
A+2\sqrt{\zeta A(\zeta A+\eta)}}{\eta+2\zeta A_p+2\sqrt{\zeta
A_p(\zeta A_p+\eta)}}\Bigg|\Bigg]
\end{equation}
where, $$\mu= \frac{b}{8{\beta}^{2}},\quad\quad
\eta=-\pi{\alpha}^{2} l_{p}^{2}(\gamma+1), \quad \quad \zeta=
1-{\alpha}^{2}{\beta}^{2},$$
\begin{equation}
\label{math:3.12} A_p=\frac{\pi\alpha^2
l_p^2(1+\gamma)}{(1-\alpha^2\beta^2)}
\end{equation}
This is the most general form of the black hole entropy which can be
obtained from perturbational approach based on GUP.\\
Expansion of (17) gives
\begin{equation}
\label{math:3.13} S\simeq\sum_{n=1}^\infty D_n (A-A_p)^n.
\end{equation}
The coefficients of this expansion have very complicated form. The
first coefficient is
\begin{equation}
\label{math:3.22} D_1=\mu\Bigg(\frac{-\frac{2\zeta
A_p+\eta}{\sqrt{A_p(\zeta A_p+\eta)}}+\zeta+1}{-2\sqrt{A_p(\zeta
A_p+\eta)}+A_p(\zeta+1)+\eta}+\sqrt{\zeta}\frac{\frac{2\zeta^2A_p+\zeta
\eta}{\sqrt{\zeta A_p(\zeta A_p+\eta)}}+2\zeta}{2\sqrt{\zeta
A_p(\zeta A_p+\eta)}+2\zeta A_p+\eta}\Bigg).
\end{equation}
The matter which is important in our calculations is the fact that
expansion (19) has no logarithmic term. In other words, since
expansion (19) contains only integer power of $A-A_{p}$, we conclude
that in microcanonical ensemble, there is no logarithmic corrections
due to quantum gravitational effects for thermodynamics of black
holes. Adler {\it et al} have found vanishing entropy for remnant in
their paper[21]. In other words, their result for entropy vanishes
when one considers Planck mass limit. In our framework, when $A =
A_p$, one finds $S=0$ and therefore remnant has zero entropy. A
result which physically can be acceptable since small classical
fluctuations are not allowed at remnant scales because of the
existence of the minimum length.

\section{Summary}
In this paper, using generalized uncertainty principle in its most
general form as our primary input, we have calculated microcanonical
entropy of a black hole. We have shown that in perturbational
expansion there is no logarithmic pre-factor, which has been the
source of controversies in literatures. Actually in calculation of
entropy we should compute the number of possible microstates of the
system and there are two possible choices for corresponding
ensemble: canonical and microcanonical ensemble. We have shown that
the contribution of microcanonical ensemble itself is vanishing. If
there is any contribution related to canonical ensemble, it cannot
cancel vanishing contribution of microcanonical ones. This argument
resolves part of controversies regarding mutual cancelation of two
contributions as have been indicated in introduction. \\

\end{document}